# D retention and material defects probed using Raman microscopy in JET limiter samples and beryllium-based synthesized samples.

C. Pardanaud[a*], M. Kumar[a], P. Roubin[a], C. Martin[a], Y. Ferro[a], J. Denis[a], A. Widdowson[b], D. Douai[c], M. J. Baldwin[d], A. Založnik[d], C. Lungu[e], C. Porosnicu[e], P. Dinca[e], T. Dittmar[f], A. Hakola[g], and EUROfusion WP PFC contributors[h] and JET contributors

[a] Aix Marseille Univ, CNRS, PIIM UMR 7345, 13397, Marseille, France
[b] United Kingdom Atomic Energy Authority, Culham Centre for Fusion Energy, Abingdon, OX14 3DB, UK
[c] CEA, IRFM, F-13108, Saint-Paul-Lez-Durance, France
[d] Center for Energy Research, University of California at San Diego, La Jolla CA, USA
[e] National Institute for Laser, Plasma and Radiation Physics, Magurele, Bucharest, Romania
[f] Forschungszentrum Jülich GmbH, Institut für Energie- und Klimaforschung - Plasmaphysik, 52428 Juelich, Deutschland, Germany
[g] VTT, P. O. Box 1000, 02044 VTT, Finland
[h] See the author list in "S. Brezinsek et al 2017 Nucl. Fusion 57 116041"

*Corresponding author: tel.: +33 622146600, e-mail: cedric.pardanaud@univ-amu.fr

## Abstract

We report on the detection by means of Raman spectroscopy of amorphous beryllium deuteride, $BeD_2$, in magnetron sputtered deposits synthesized in two different laboratories and containing about 20 at.% of deuterium. In contrast, this signature has not been found for the JET limiter samples studied coming from the inner, outer or upper limiters, even when coming from a deposition zone of the limiters. We give a way to disentangle that $BeD_2$ signature from other signatures falling in the same spectroscopic range and mainly related to other phenomena. We also analyze the Raman characteristics of the JET sample defects. These results could help in the interpretation of D thermal desorption spectra and in future analyses of JET thick Be deposit divertor tiles.

# 1. Introduction

Tritium accumulation is an issue for ITER and consequently processes involving T trapping need to be investigated. Retention and release in ITER first wall co-deposits have to be estimated using knowledge acquired from currently operating tokamaks and from laboratory experiments [1-4]. In the Joint European Torus (JET) tokamak in the ITER-like wall (ILW) configuration (divertor in tungsten, W, and limiters in beryllium, Be), a hydrogen isotope inventory has been recently performed on limiters during the three first campaigns, 2011-2012, 2013-2014 and 2015-2016, labelled respectively ILW1, ILW2 and ILW3 [5-7]. It has been found that, as in the past with other tokamaks [8], the deuterium (D) content is low in components exposed to high heat flux and in erosion zones, and high in deposition zones. However, the D/Be amount is one order of magnitude lower than of D/C, C being carbon, in the carbon era tokamaks. Thermal desorption spectroscopy (TDS) has shown that hydrogen isotopes release starts at 500 K, with two main contributing thermal ranges, 700–760 K and 850–900 K [6]. In the case of laboratory samples, a sharp D release peak is often observed at 450-460 K. This is explained by blister peeling off in the case of ion-beam implanted samples [9] (and references therein) while in the magnetron sputtering codeposited layers, another mechanism involving beryllium deuteride ($BeD_2$) formation and decomposition is invoked [10]. $BeD_2$ was observed in plasma-exposed samples using a surface technique, X-ray Photoelectron Spectroscopy (XPS) in 2009 [11] but, to our knowledge, there was no further direct observation since then, except when we reported on the growth of crystalline $BeD_2$, coexisting with blisters, at an implantation fluence of about $1\times10^{21}$ $D/m^2$, close to the saturation value [12]. It should be noted that the presence of impurities most probably play a role in the shape of TDS spectra, especially in the case of JET samples [13, 14].

Raman microscopy, and specifically Raman microscopy applied to beryllium based materials, is a technique which is not used as a routine technique as it is still in development. This work participates to this. Basically, Raman microscopy has demonstrated in the past its ability to investigate the chemistry related to beryllium, by detecting frequencies of typical vibrational modes (Be-C, Be-H), as well as the amount of defects at the microscale [15-20], and by detecting changes in the spectrum related to Be phonons (normal vibrational modes). Basically, phonons are coupled to their environment. Due to that, the spectroscopic parameters (position σ, width Γ and intensity A) of their corresponding bands could contain the information of these couplings. All that interactions are piloted by physico-chemistry which is quantum rule based. Structure of the material, and its symmetry, is an important part that leads to intensity selection rules (i.e. bands that can be seen or not seen in the spectrum). Briefly, at room temperature, Be belongs to a hexagonal-close-pack crystal lattice structure. Only one of the six normal vibrational modes is Raman active, with an irreducible $E_{2G}$ representation. When the crystallite is no more infinite, due to different kind of defects, selection rules are partially changed. It leads to the rise of some new bands. Depending on the material, that bands could be the Phonon Density of States (PDOS)bands ([21, 22], as non-exhaustive examples of materials), or other features induced by other mechanisms [23]. We present here an extended Raman microscopy analysis centered in the 300 – 1400 $cm^{-1}$ spectral range, aimed at unravelling the congestion of this spectral range to identify the signature of $BeD_2$ and to clarify that of defects. We investigate laboratory samples corresponding to a large variety of textures, defects and chemistry, present in magnetron sputtered codeposited Be layers (D alone or with helium-He, oxygen-O, nitrogen-N and neon-Ne) produced at various temperatures and a Be sample implanted at various fluences. We present the comparison with several tokamak samples exposed in JET from 2011 to 2016 and extracted from the upper, inner and outer limiters of the ILW.

2. Experimental details

Magnetron sputtering (MS) codeposited layers were synthesized at INFLPR (Bucharest), from 2015 to 2020, 3 of them with a high-D amount (≈20 at.%) and the rest with a low-D amount (≈ 0-2 at.%). They were ≈5 µm thick, deposited on W substrates by sputtering from magnetron targets bombarded with Ar using 20 sccm gas flow and different layer compositions were obtained by varying $D_2$, $N_2$, Ne, He or $O_2$ gas flows (see previous publications for more details [24]). The substrate temperature was varied, at Room Temperature (RT), 100°C and 400°C. One of the INFLPR samples reported here was obtained using the Thermal Vacuum Arc (TVA) method (see [15] for details). D-content was obtained by TOF-ERDA and NRA measurements by WPPFC contributors [19]. Two MS codeposited layers were from PISCES (San Diego) (details in [25]), intentionally produced to study the sharp D release associated with beryllium deuteride [10]. One has a high-D amount (≈21 at.%) and the second is the same sample after being heated at 673 K under UHV, thus with a low-D amount (≈4 at.%). The D-implanted sample was a polycrystalline Be sample implanted in the ARTOSS set-up with 2 keV D ions at fluence in the range 0.4-4 $10^{22}$ $D/m^2$ (details in [26]).

The properties of the seven JET samples analyzed here are summarized in Table 1 with their ID, location and type (erosion, deposition or melted), that could also be seen in figure 1.

| Location | Campaign | | ID | Type |
|---|---|---|---|---|
| Ex-vessel | Non-exposed | | TT7 | As received |
| 4D14 OPL (R5C1) | ILW2 | | 363 | Left deposition zone |
| 4D14 OPL (R5C22) | ILW3 | | 544 | Right deposition zone |
| 2XR10 IWGL (R5C10) | ILW1 | | 053 | Erosion zone |
| 4D14 OPL (R4C10) | ILW1 | | 131 | Erosion zone |
| 3A8 DP (R3C6) | ILW3 | | 414 | Tile ridge-partially melted |
| 2B2C DP (R4C5) | ILW3 | | 450 | Above the tile ridge |

*Table 1.* List of the analyzed JET samples with their ID and location at the JET-ILW vacuum vessel. OPL= Outer Poloidal Limiter. DP=Dump Plate. IWGL=Inner Wall Guard Limiter.

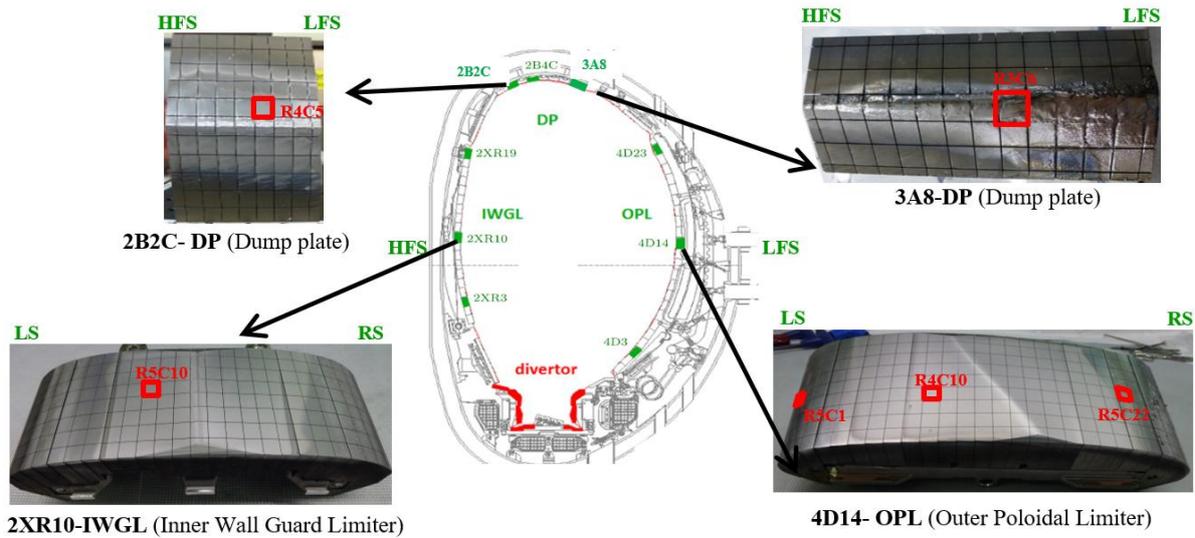

*Figure 1. Scheme of the JET cross-section with the locations and the photographs taken after the ILW1 campaign of the dump plate tiles (2B2C-DP and 3A8-DP), the inner wall guard limiter tile (2XR10-IWGL) and the outer poloidal limiter tile (4D14-OPL). Red marks show the location of analyzed samples. The high field side (HFS) and low field side (LFS) (the inner and outer sides, respectively) are indicated for the upper limiter tiles (DP tiles). For the inner and outer limiter tiles, left and right sides (LS and RS) are the toroidal positions when facing the components from inside the vacuum vessel. Dark zones at the left and the right of the IWGL and OPL tiles are deposition zones. The center of these tiles are erosion zones.*

Raman spectra were recorded in the standard back-scattering geometry using a Horiba-Jobin-Yvon HR LabRAM HR800 Raman microspectrometer, with $\lambda_L$ = 514.5 nm, × 100 magnification and numerical aperture of 0.9 (i.e. a laser spot radius of ≈ 0.34 µm). We used the Raman mapping mode with the XY motorized stage to check the spatial inhomogeneity at the micrometer scale or conversely, to get an averaged spectrum on a zone larger than the laser spot. Raman mapping area ranged from 25 to 6000 µm$^2$ with steps of typically a few micrometres, which correspond to running times from a few hours to four days, a single spectrum acquisition time being typically 100 - 1000 s. A description, and a scheme of the set-up could be found in [27]. For JET samples, as we observed inhomogeneity at the centimetre scale (for example from a melted to a non-melted zone), we recorded mappings at their corners and at their centre, while for laboratory samples, as they were more homogeneous, we recorded mappings only at their centre.

Spectra of crystalline materials display thin Raman bands sometimes accompanied with weak and broad satellites due to a large number of transitions becoming Raman active because of defects or disorder and whose spectral density is given by the PDOS. The main Raman band of crystalline Be, the $E_{2G}$ band, lies at 455 cm$^{-1}$ [28]. As we were concerned here by disorder and by its PDOS Raman signature, we plotted the PDOS alone after fitting the 350-650 cm$^{-1}$ spectral range by 6 symmetric bands (5 are due to Gaussian contributions in the PDOS and one, more intense and mainly Lorentzian, due to the $E_{2G}$ mode) and subtracting the band corresponding to the $E_{2G}$ mode ~~band~~.

3. **Results**

*3.1 Beryllium deuteride identification in laboratory samples*

The weak PDOS signal due to the presence of defects is detected in the 350-650 cm$^{-1}$ region, around the main crystalline Be band at 455 cm$^{-1}$, ($E_{2G}$ vibrational mode). A second signal is detected in the 750 – 1300 cm$^{-1}$ region, a frequency range double of that of PDOS, and that we call for this reason 2PDOS [16]. This 2PDOS feature may therefore be the second harmonic of the PDOS. However, it should not be as intense as it is. Some coupling mechanism like the one described in [29] could be at the origin of its intensity enhancement. This is not the purpose of this work to understand this point, but we have to take into account the existence of this spectroscopic signature as it falls close to other spectroscopic signature of interest that we are searching for.

Fig. 2-a displays several Raman spectra in the 300-1400 cm$^{-1}$ region to emphasize the possible other contributions close to this PDOS-2PDOS spectral range. The upper spectrum concerns reference amorphous $BeD_2$ (a-$BeD_2$). There are several studies reporting IR spectra of various beryllium hydrides and deuterides under molecular forms, either in the gas phase, either isolated in cryogenic matrices [30-32]. These studies give information on the positions of the IR active vibrational modes, i.e. the antisymmetric stretching and the bending modes. In [32], the authors mention the position of the three vibrational modes (i.e with the symmetric stretching mode) in the case of the a-$BeH_2$ and a-$BeD_2$. There exists only one reported Raman spectrum of the amorphous hydride (a-$BeH_2$): it displays a main broad band centered at about 1400 cm$^{-1}$ corresponding to the Raman active symmetric stretching mode [33]. In the plot of Fig. 2-a, we have reported this spectrum by applying a coefficient for the wavenumbers to take into account the isotopic shift (H to D), using the square root of the ratio of the reduced masses Be/D and Be/H (~1.35). This coefficient is consistent with those estimated from the various experimental values reported for beryllium hydrides and deuterides, found in the range 1.29 – 1.40 [30-32]. Fig.2-a shows that the $BeD_2$ spectrum then falls in the 2PDOS region. Fig.2-b also shows that various beryllium oxide signatures could also fall in the region studied, meaning there is an overlap. We display here the reference BeO PDOS that is in the range 650-750 cm$^{-1}$ [34] while the main thin band corresponding the wurtzite crystalline phase peaks at ~700 cm$^{-1}$. We also display the PDOS of a more exotic phase, $BeO_2$, that could introduce modes with wavenumbers as high as 1125 cm$^{-1}$ [35]. The so-called 2PDOS region is thus crowded and could contain Raman signatures of various origins such as beryllium deuteride, beryllium oxide or defective beryllium.

Fig.2-b shows the 300-1400 cm$^{-1}$ region for 5 representative laboratory samples, high-D (in blue, 2 spectra) and low-D (in red) MS samples from INFLPR and high-D (in green) and low-D (in magenta) MS samples from PISCES. In the PDOS region, bands are sharper for low-D than for high-D samples. Consistently, the spectral decomposition shows narrower and less overlapping components for low-D than for high-D samples. Such a line broadening is expected when disorder increases. In the 2PDOS region, two very broad bands at ~800 and ~1000 cm$^{-1}$, probably including several components, dominates the 700 – 1100 cm$^{-1}$ region. The relative intensity between these two peaks is clearly inverted from low-D to high-D samples and the peak at 1000 cm$^{-1}$, where the main a-$BeD_2$ signature is supposed to lie, is larger for the latter. For the low-D samples, two small peaks at 1150 and 1290 cm$^{-1}$ are observed [16], and will be investigated in another publication. For the high-D samples, an extra band at 1200 cm$^{-1}$ is observed, that may be interpreted to a-$BeD_2$ dangling bond or to the presence of an unknown oxide. Fig. 2-c displays the integrated area of the 2PDOS region as a function of the integrated area of the PDOS region after normalization to the $E_{2G}$ peak ($A_{2PDOS}/A_{E2G}$ versus $A_{PDOS}/A_{E2G}$) for all the samples studied. $A_{PDOS}/A_{E2G}$ gives an indication of the amount of defects: the higher the normalized PDOS, the more defects in the sample. The data points follow a straight line except for the high-D MS samples (although in a lower extent for the PISCES sample). This 2PDOS intensity increase corresponds to the a-BeD$_2$ peak at 1000 cm$^{-1}$ observed in Fig.2-a. Fig.2-d displays the integrated area of the 2PDOS region divided by that of the PDOS ($A_{2PDOS}/A_{PDOS}$) versus the normalized integrated area of the PDOS region ($A_{PDOS}/A_{E2G}$), showing the decrease of the relative intensity of the two regions from 3 to 0.2.

To summarize the findings about figure 2:
- We detected that the intensity related to the PDOS feature is continuously related to the intensity of the 2PDOS feature for most of the samples used. It is not related specifically to the presence of D.
- The 2PDOS intensity vary relatively to the PDOS intensity. It is not related specifically to the presence of D. The physical origin of this relation is not under the scope of this work.
- The presence of the 2PDOS feature impedes BeD$_2$ signature detection for low amount of D, but despite 2PDOS feature presence, it is possible to detect BeD$_2$ for high amount of D.
- No BeD$_2$ signatures were detected in JET limiter samples

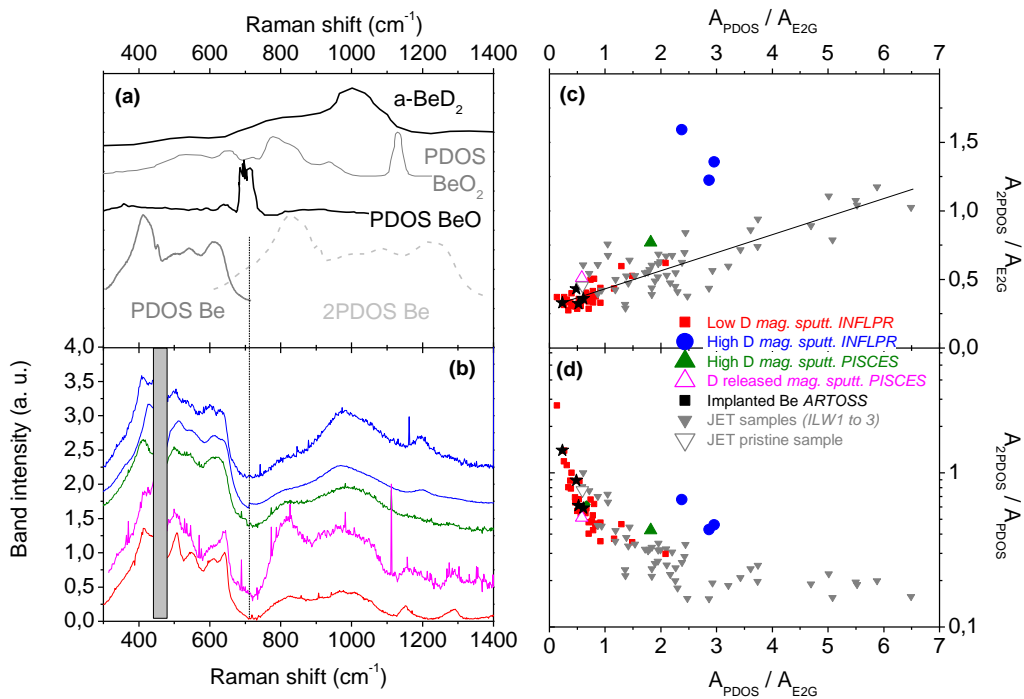

*Figure 2. Raman Be PDOS and 2PDOS signals of Be-based samples: (a) Reference spectra (b) Raman spectra recorded for selected high-D and low-D MS samples. (c) $A_{2PDOS}/A_{PDOS}$ versus $A_{PDOS}/A_{E2G}$. (d) $A_{2PDOS}/A_{E2G}$ versus $A_{PDOS}/A_{E2G}$. Laboratory samples data points are colored and JET data points are in grey. Color code in figure b is the same as for figure c and d.*

## 3.2 Defect characterization in JET samples

In Fig. 3 we probe here defects with the normalized PDOS intensity, $A_{PDOS}/A_{E2G}$, the $E_{2G}$ band width, $\Gamma_{E2G}$, and the $E_{2G}$ band position, $\sigma_{E2G}$. Fig. 3-a shows that the JET samples cover a large zone of $A_{PDOS}/A_{E2G}$ (0 – 7) and $\Gamma_{E2G}$ (10 – 25 cm$^{-1}$), contrary to the MS laboratory samples that are grouped at low values of $A_{PDOS}$ (0 -1) and low values of $\Gamma_{E2G}$ (7-11 cm$^{-1}$). The non exposed as received JET sample behaves as lab samples. The three types of JET samples (eroded, deposited and melted) are found very defective and very heterogeneous. By comparison, despite their large variety of synthesis conditions, all the laboratory samples are found comparatively less defective and less heterogeneous/more homogeneous. JET eroded samples are found the least heterogeneous of the JET samples. Their data points are in two zones: one mainly vertical, centered at $\Gamma_{E2G}$=11 cm$^{-1}$, and one mainly horizontal, around $A_{PDOS}/A_{E2G}$ =1, which is close to data points of implanted sample. JET deposited samples are found the most defective and heterogeneous, with a large vertical dispersion centered at $\Gamma_{E2G}$=15 cm$^{-1}$. JET melted samples are the most heterogeneous, with data points close to both those eroded and deposited. Fig. 3-b shows that the frequencies fall mainly in the range $\sigma_{E2G}$=450-455 cm$^{-1}$, whatever the sample origin, except for part of the eroded samples which are found at 455-460 cm$^{-1}$ (in majority sample 053). This frequency upshift may be due to a compressive stress induced by the presence of impurities. Finally, Fig. 3-c displays the $\sigma_{E2G}$ evolution under the Raman laser exposition when varying the laser power for the two PISCES samples. It shows that $\sigma_{E2G}$ varies in a 5 cm$^{-1}$ range, decreasing when heating, in a less extend for the pre-heated sample. This evolution (reversible) points out the possible influence of the sample thermal behavior during Raman measurements in the measured values of $\sigma_{E2G}$, like it was done for tungsten oxides in [36].

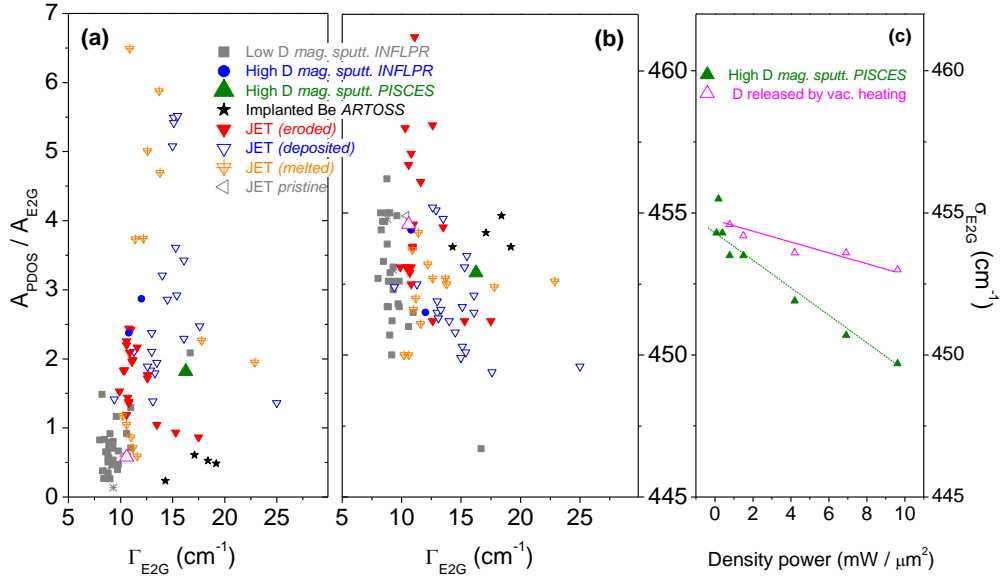

***Figure 3.*** *Comparison of laboratory Be-based and JET samples: (a) $A_{PDOS}/A_{E2G}$ versus $\Gamma_{E2G}$. (b) $\sigma_{E2G}$ versus $\Gamma_{E2G}$. (c) $\sigma_{E2G}$ versus laser density power for PISCES samples. JET data points are colored and laboratory samples data points are in grey.*

## 4. Concluding remarks

We have studied various Be-based laboratory samples and compared them to JET limiter samples from erosion, deposition and melted zones. We have identified the Raman signature of a-BeD$_2$ in magnetron sputtered deposits containing 20 at.% D. No a-BeD$_2$ has been detected for JET samples, even in deposition zones, probably due to its low decomposition temperature (540 K [16]). There are other possibilities for D trapping, not addressed in this paper, such as D in vacancies or bubbles, D under D$_2$ molecules, or D bonded with impurities such as O and C. We have also studied the Raman signature of defects and found that the JET samples were very defective and very heterogeneous, without a clear correlation between defect signature and their eroded, deposited or melted types. Laboratory samples were found much less defective than JET samples, and despite their large variety of synthesis conditions, they are also much less heterogeneous. This work highlights the capability of laboratory samples to address particular physical or chemical questions of interest for ITER (here the identification of BeD$_2$), but also the difficulty to directly compare them with tokamak samples. This is not very surprising considering the hard and complex conditions undertaken by the latter. Care must be taken in Raman interpretations for various reasons. It is a microprobe and then it is difficult and time consuming to probe large and heterogeneous samples. In addition, materials could be sensitive to the laser heating, preventing from fast acquisition time and possibly leading to misinterpretation. There is also possible misinterpretation due to spectral similarities: a previous attribution of D bonded to O in Be deuteroxides [17] is nowadays questioned as there exist fatty acids with very similar signatures. This study shows that Raman microscopy could probably be a useful technique to check for a-BeD2 in JET divertor tiles.
Note that no sample from JET expected to have a significant amount of deuterium have yet been investigated using this technique.

**Acknowledgements**

This work has been carried out within the framework of the EUROfusion Consortium and has received funding from the Euratom research and training programme 2014–2018 and 2019–2020 under grant agreement No. 633053. The views and opinions expressed herein do not necessarily reflect those of the European Commission. Work performed under EUROfusion WP PFC.